\documentclass[useAMS,usenatbib]{biom}
\usepackage{amsfonts,amsmath, amssymb}

\def\ind{\perp\!\!\!\perp}
\def\notind{\not\!\perp\!\!\!\perp}

\newcommand{\Pb}{\mathbb{P}}
\newcommand{\Pn}{\mathbb{P}_n}
\newcommand{\E}{\mathbb{E}}

\DeclareSymbolFont{bbold}{U}{bbold}{m}{n}
\DeclareSymbolFontAlphabet{\mathbbold}{bbold}

\usepackage[figuresright]{rotating}

\title[Discussion]{Discussion of ``Data-driven confounder selection via Markov and Bayesian networks'' by Jenny H{\"a}ggstr{\"o}m}

\author{Edward H. Kennedy$^{*}$
 and
Sivaraman Balakrishnan$^{\dagger}$ \\
$^*$\emph{email:} edward@stat.cmu.edu \\
$^{\dagger}$\emph{email}: siva@stat.cmu.edu\\
Department of Statistics, Carnegie Mellon University, Pittsburgh, PA 15213, U.S.A. \\
}

\begin{document}

\date{{\it Received XXX} XXXX. {\it Revised XXX} XXXX.\newline 
{\it Accepted XXX} XXXX.}

\pagerange{\pageref{firstpage}--\pageref{lastpage}} \pubyear{XXXX}

\volume{XX}
\artmonth{XXX}
\doi{XXXX}

%  This label and the label ``lastpage'' are used by the \pagerange
%  command above to give the page range for the article

\label{firstpage}

%  pub the summary here
%In this discussion we consider why it is important to estimate causal effect parameters well even they are not identified, propose a partially identified approach for causal inference in the presence of colliders, point out an under-appreciated advantage of double robustness, discuss the relative difficulty of independence testing versus regression, and finally commend H{\"a}ggstr{\"o}m for her exploration of causal inference with high-dimensional confounding, and make a call for further research in this same vein.

%
%  Please place your key words in alphabetical order, separated
%  by semicolons, with the first letter of the first word capitalized,
%  and a period at the end of the list.
%


\maketitle

We congratulate Jenny H{\"a}ggstr{\"o}m on an interesting article exploring the use of graph estimation to aid confounder selection, and
we appreciate the opportunity to comment.

\section{Questioning Unconfoundedness \& An Answer}
In most observational studies the assumption of unconfoundedness given all measured covariates (i.e., that $Y(t) \ind T \mid X$) is very unlikely to hold, let alone unconfoundedness given a smaller subset (i.e., that $Y(t) \ind T \mid S$ for some $S \subset X$). Both assumptions seem to be required for H{\"a}ggstr{\"o}m's proposed approach to be effective. So, although we appreciate the author and others' explorations of how to optimally select adjustment covariates for efficiency purposes, we hope that in doing so the forest is not lost for the trees. In particular, for most observational studies the observed data parameter
$$ \beta = \E\{ \E(Y \mid X, T=1) - \E(Y \mid X, T=0) \} $$
is not equal to the the causal parameter 
$$ \beta^* = \E\{Y(1)-Y(0)\} $$
due to unmeasured confounding (i.e., $Y(t) \notind T \mid X$). In our view, H{\"a}ggstr{\"o}m's paper is mostly about reducing the dimension of $X$ so as to estimate $\beta$ more efficiently. 

One might then wonder: is it really so useful to put forth so much effort in estimating $\beta$ well, when it is typically not even equal to the causal effect $\beta^*$ that we actually care about? 

In fact, we believe the effort is worth it. The reason is that the observed data parameter $\beta$ still plays a critical role in bounding problems and sensitivity analyses that do not assume unconfoundedness, i.e., even when $\beta \neq \beta^*$. Specifically, without assuming unconfoundedness, we can write the causal effect as (for example)
$$ \beta^* = \beta - \E\{ \gamma(X,1-T) \} $$
%$$ \beta - \alpha \leq \beta^* \leq \beta + \alpha , $$
where the bias function
$$ \gamma(X,t) = \E\{ Y(t) \mid X, T=t\} - \E\{Y(t) \mid X, T=1-t\} $$
%$$ \alpha \geq \max_t | \E\{Y(t) \mid X,T=t\} - \E\{Y(t) \mid X, T=1-t\} | $$
captures the extent of unmeasured confounding for estimating $\E\{Y(t)\}$ among those with covariates $X=x$. (Note that in the special case of unconfoundedness, i.e., $Y(t) \ind T \mid X$, we have $\gamma(x,t)=0$).

Hence, even if the bias function $\gamma$ is only known up to its sign or is only known to be bounded (or bounds are specified and varied in a sensitivity analysis), it is still essential to estimate the purely observed data parameter $\beta$ well, since $\beta^*$ is a function of $\beta$. In other words, approaches  that focus on statistical issues of estimating $\beta$ well (such as H{\"a}ggstr{\"o}m's) remain critically important, even if there is unmeasured confounding and $\beta \neq \beta^*$. If we estimate parameters like $\beta$ with bias or at slow rates of convergence, then we will have the same problems in constructing bounds and doing sensitivity analyses (simply because they also typically require estimating $\beta$). 

We would be curious to hear the author's thoughts about whether her work yields any additional benefits in sensitivity analyses or other settings where unconfoundedness assumptions are weakened. This is related to our next point.

\section{The Collider Problem}

The backdoor path criterion \citep{pearl2009causality} allows one to determine graphically whether conditioning on a given covariate set $S \subseteq X$ ensures unconfoundedness, i.e., whether $Y(t) \ind T \mid S$. In order to yield identification unconfoundedness must hold given some subset $S$, but the backdoor path criterion does not require that unconfoundedness holds given the full covariate set $X$ (i.e., we do not need $Y(t) \ind T \mid X$). However, the backdoor path criterion requires knowledge of the entire causal diagram, including relationships among covariates, which is often not available.

Alternatively, \citet{de2011covariate} considered covariate selection without knowledge of the entire causal graph, but requiring unconfoundedness given the full covariate set $X$. Note that this rules out the presence of colliders, which for our purposes can be defined as variables $C$ linked to treatment and outcome via paths like $A \leftarrow U_1 \rightarrow C \leftarrow U_2 \rightarrow Y$. Adjusting for colliders yields so-called M-bias resulting from the path being unblocked by conditioning \citep{pearl2009causality}. 

\citet{vanderweele2011new}, on the other hand, proposed a selection criterion that requires neither full knowledge of the causal graph, nor unconfoundedness given the full covariate set (and so allows for the presence of colliders). Their criterion simply says to adjust for all pretreatment variables that are a cause of either treatment or outcome. However, this means one must know which variables are colliders (or otherwise not a cause of treatment or outcome).

In our view, H{\"a}ggstr{\"o}m's approach seems to be an alternative to \citet{de2011covariate} that operates under the same assumptions. For example, the theoretical results require strong unconfoundedness assumptions (ruling out not only unmeasured confounders, but also any unmeasured variables), and the methods fail when $Y(t) \notind T \mid X$ in Simulation Setting 2, even though unconfoundedness holds given a subset.

Summarizing, suppose
\begin{enumerate}
\item[(A0)] there exists a subset $S \subseteq X$ such that $Y(t) \ind T \mid S$,
\end{enumerate}
and consider the following further conditions:
\begin{enumerate}
\item[(A1)] the full causal graph is known,
\item[(A2)] $Y(t) \ind T \mid X$ (implying there are no colliders),
\item[(A3)] it is known which variables are colliders.
\end{enumerate}
Then, in addition to A0, the backdoor path criterion requires A1, the \citet{de2011covariate} and H{\"a}ggstr{\"o}m approach requires A2 (but not A1), and the \citet{vanderweele2011new} approach requires A3 (but not A1 or A2). This begs the question: what should we do when A0 holds, but not A1, A2, or A3? In other words, what if unconfoundedness holds given a subset $S$, but there may be colliders, and unfortunately we do not know where they are?

One possibility is to consider inference under A0 and the weaker assumption (relative to A1--A3) that
\begin{enumerate}
\item[(A4)] there are fewer than $k$ colliders.
\end{enumerate}
For example, suppose it is known that there are no more than $k=1$ colliders, but it is unknown which if any variable might be the collider. Then, letting $X_{-j} = X \setminus X_j$ denote the covariate set excluding covariate $j$ (with $X_{-0}=X$), it must be that at least one of the leave-one-out parameters
$$ \beta_j = \E\{\E(Y \mid X_{-j}, A=1) - \E(Y \mid X_{-j},A=0)\} $$
equals the true causal effect $\beta$, for $j \in  \mathcal{J}=\{0,...,|X|\}$. Hence the true effect $\beta$ is partially identified by the set of values $\{\beta_j: j \in  \mathcal{J}\}$ and is bounded by the range  $[\min_j \beta_j, \max_j \beta_j]$. Further, in cases where certain covariates are known to not be colliders we can incorporate this information. 
It seems likely that in some problems these bounds could be quite narrow and thus informative, even in the presence of unknown colliders. %Effectively these bounds give a quantitative sense for the bias that might be incurred by conditioning on (unknown) colliders.

Note that for general $k>1$ one could use the same approach, except letting $j$ be a multi-index spanning all combinations of $k$ or fewer covariates. For  $k=2$, for example, we would have $j=(j_1,j_2) \in \mathcal{J}^2$ with $j_1 < j_2$. 
%Further, just as the \citet{de2011covariate} and H{\"a}ggstr{\"o}m approach could be used after that of \citet{vanderweele2011new} to find a minimal subset $S \subset X$, one could imagine doing so 
It would be useful to determine whether there may be other better approaches for dealing with unknown collider structure.

\section{Inference After Data-Driven Selection}

In this section we consider H{\"a}ggstr{\"o}m's use of propensity score matching versus targeted maximum likelihood estimation (TMLE), and make the case that doubly robust influence-function-based approaches are uniquely effective (if not necessary) in realistic settings where covariate adjustment requires some flexible data-adaptive modeling, e.g., via  high-dimensional nonparametric regression methods.

Doubly robust influence-function-based estimators of $\beta$ (such as TMLE \citep{van2011targeted}) generally take the form
$$ \widehat\beta_{\text{dr}} = \Pn\left[ \frac{(2T-1)\{Y-\widehat\mu_T(X)\}}{(2T-1) \widehat\pi(X) + (1-T)} + \widehat\mu_1(X)-\widehat\mu_0(X) \right] $$
where $\Pn\{f(X)\}=\frac{1}{n} \sum_i f(X_i)$ denotes sample averages, $\widehat\mu_T(X)$ is an estimate of $\E(Y \mid X,T)$, and $\widehat\pi(X)$ is an estimate of the propensity score $\Pb(T=1 \mid X)$. For more details we refer to \citet{bickel1993efficient},  \citet{van2003unified}, \citet{bang2005doubly}, and \citet{tsiatis2006semiparametric}.

It is commonly claimed that the benefit of doubly robust estimators is that they give two chances at consistency and asymptotic normality, as long as either $\pi$ or $\mu_t$ is estimated with a correct parametric model (not necessarily both). While this is true, relying on one of two parametric models is essentially as risky as relying on a single parametric model, simply because most parametric models are probably wrong. 

In our opinion, the crucial virtue of doubly robust estimators (and influence-function-based estimators in general) is that they can converge at fast parametric rates to the true $\beta$ (and yield nice centered Gaussian limiting distributions), even when the functions $\pi$ and $\mu_t$ are estimated nonparametrically at slower rates, e.g., via flexible regression methods. This is not true of other general estimators, such as those based on propensity score matching or regression \citep{van2014higher}. As soon as one moves beyond the world of parametric models, the behavior of these estimators is immediately degraded, resulting in slower convergence rates and very limited  possibility of constructing tight confidence intervals. %In other words, non-influence-function-based estimators that are not specifically tailored to do a good job in estimating their target parameter typically inherit the poor properties of their nuisance estimators (like $\hat\pi$ and $\hat\mu$).

This under-appreciated aspect of double robustness is a result of the fact that, under empirical process conditions (which can be avoided via sample splitting), we have
$$ \frac{\widehat\beta_{\text{dr}} - \beta}{\sigma/\sqrt{n}} = Z + \sqrt{n} R_2 + o_\Pb(1)  $$
for $Z \sim N(0,1)$ a standard Gaussian random variable, $\sigma$ the asymptotic standard deviation, and $R_2$ a ``second-order'' remainder term with
$$ |R_2| \lesssim \max_{t} \| \widehat\mu_t - \mu_t \| \cdot \| \widehat\pi - \pi \| $$
where $\| f \|^2 = \int f(x)^2 d\Pb(x)$ is the squared $L_2(\Pb)$ norm. Therefore even if $\|\widehat\pi-\pi\|$ and $\|\widehat\mu_t-\mu_t\|$ converge at slower than $\sqrt{n}$ rates, $\widehat\beta_{\text{dr}}$ can still converge at a fast $\sqrt{n}$ rate as long as the product of the $\widehat\pi$ and $\widehat\mu$ rates are faster than $\sqrt{n}$. For example, this will occur if $\|\widehat\pi-\pi\| \asymp \| \widehat\mu_t-\mu_t\| = o_\Pb(n^{-1/4})$, which is a rate one could plausibly attain under sparsity, smoothness, or other nonparametric structural constraints.

This is important in the current setting since the stepwise graph estimation approaches used by H{\"a}ggstr{\"o}m to aid variable selection seem to be highly non-smooth procedures that would in general yield estimates of $\pi$ and $\mu_t$ that converge at slower than parametric rates. Hence we conjecture that using estimators not based on influence functions will in general yield slow rates and intractable inference. But using doubly robust estimators (with sample splitting) allows for fast rates and valid inference under relatively weak rate requirements. It would be interesting to explore under what conditions such rates might be attainable for  H{\"a}ggstr{\"o}m's proposed approach. Accordingly, we consider this problem in the next section. 

\section{Independence Testing Versus Regression}

In this paper H{\"a}ggstr{\"o}m proposes estimating the graph (i.e., independence relationships between variables), reading off a reduced adjustment covariate set from the estimated graph based on various criteria, and then applying usual adjustment techniques (propensity score matching based on parametric propensity score modeling, and TMLE using BART regression) to this reduced covariate set. She compares this approach via simulation to alternatives that use Random Forests and Lasso to reduce the covariate set, as well as applying usual adjustment methods based on the full covariate set. 

This raises the question of how the above approaches would compare to using high-dimensional regression methods (such as Random Forests or Lasso) to estimate the nuisance functions $\pi$ and $\mu_t$ directly, rather than only using them to select a reduced adjustment set. More generally one wonders under what conditions using independence testing to reduce the dimension of $X$ and then applying regression methods achieves better performance than simply applying high-dimensional regression methods directly, say in terms of $L_2(\Pb)$ error
$$ \| \widehat\mu_t - \mu_t \| = \sqrt{\int \Big\{ \widehat\mu_t(x) - \mu_t(x) \Big\}^2 \ d\Pb(x)} , $$
since as we saw in the previous section this is largely what matters for obtaining high-quality estimates of $\beta$ that converge at fast rates.

In fact the author's simulation results (e.g., Figure 4) seem to indicate that there might not be much additional gain from confounder selection when one is already using doubly robust estimators with the aforementioned second-order bias property (and flexible estimators for the nuisance functions $\pi$ and $\mu_t$). Correspondingly, it would be interesting to see how a doubly robust estimator that used the Lasso or Random Forests instead of BART might fare. One could even consider ensembles of all of these methods, with and without confounder selection, for example using cross-validation via Super Learner \citep{van2011targeted}. 

Concretely, we feel that further study is warranted to understand conditions under which one can reduce the size of the covariate set more statistically efficiently than one can directly perform nonparametric regression on the full covariate set.
For example, suppose we ignore the computational (and statistical) cost of searching for an appropriate subset of variables, and were instead to focus on two fixed subsets $V$ and $W$ with $X=(V,W)$.
In order to eliminate the subset $W$ from consideration, we need to test if 
$$ W \ind Y \mid T, V. $$
Even in the case when $V$ is low-dimensional, so that the following step of estimating the causal effect is not a statistical bottleneck, testing this independence statement involves rates that under standard nonparametric assumptions depend exponentially on the size of the covariate set $W$. While it is plausible that under certain assumptions independence testing can be statistically cheaper than just estimating the regression function, it will be important to carefully articulate and examine what exactly these assumptions might be.
Independence testing is a difficult nonparametric problem in its own right \citep{paninski2003estimation,jiao2015minimax}, and the method of testing independence to facilitate a subsequently more efficient lower-dimensional regression is unfortunately not a panacea that avoids the curse of dimensionality.

Finally we also point out that estimation of a minimal adjustment set $S \subseteq X$ might be of interest per se, for example to inform future studies and reduce data collection burden. Such a goal should be treated separate and apart from that of estimating the parameter $\beta$. In short, methods that do a good job finding a minimal adjustment set should not necessarily be used as a pre-processing step if the goal is to estimate $\beta$ well; these are two different problems with potentially very different criteria for success. 

\section{Dimension Reduction \& Minimax Efficiency}

We conclude this discussion by highlighting the importance of further exploration of treatment effect estimation with complex high-dimensional covariates, as pursued here by H{\"a}ggstr{\"o}m via an independence testing-based dimension reduction approach. We give a brief example of how the causal inference landscape can change quite drastically when one moves beyond parametric models, and hope to convince the reader that there are many important unanswered questions in this area that warrant further study.

For an example, \citet{robins1995semiparametric} and \citet{hahn2004functional} showed that the efficiency bound for estimating $\beta$ does not change if it is known that the propensity score depends on only a subset of the covariates $X$. (The former framed this in terms of the equivalent problem of characterizing efficiency under missing at random and missing completely at random assumptions). This fact shows why covariate adjustment is useful even in completely randomized trials, and has also been important in informing and analyzing confounder selection approaches, including those that  H{\"a}ggstr{\"o}m uses \citep{de2011covariate} and others \citep{white2011causal}.

However, the above efficiency bound is only relevant if $\sqrt{n}$ rates are attainable, and the story is quite different if one considers nonparametric settings where this may not be the case. For example, suppose the propensity score $\pi$ and regression functions $\mu_t$ are $d$-dimensional and lie in H{\"o}lder classes with smoothness parameters $\alpha$ and $\zeta$, respectively. %H{\"o}lder classes are widely used in nonparametric function estimation and essentially mean the functions have bounded partial derivatives up to order equal to the smoothness parameter. Therefore this would mean for the propensity score $\pi$ that all partial derivatives exist up to order $\alpha$ and are bounded (we restrict the smoothness parameters to be integer-valued for simplicity). Clearly larger values of the smoothing parameters indicate that more smoothness is assumed.
Then there exist estimators such that
$$ \| \widehat\pi - \pi \| = O_\Pb\left(n^{\frac{-\alpha}{2\alpha+d}} \right) \ , \ \| \widehat\mu_t - \mu \| = O_\Pb\left(n^{\frac{-\zeta}{2\zeta+d}} \right), $$
and based on the expression for the second-order remainder $R_2$ from Section 3, for these nuisance estimators the doubly robust estimator of $\beta$ will have rate of convergence $n^{-\xi}$ for
$$ \xi = \left( \frac{\alpha}{2\alpha+d} + \frac{\zeta}{2\zeta+d} \right) \wedge \frac{1}{2} , $$
as noted by \citet{robins2017minimax}. Clearly if it is known that the propensity score only depends on $d^* < d$ variables (assuming no change in smoothness), and if one is outside the $\sqrt{n}$ rate regime, then this knowledge can yield a faster rate of convergence. This goes to show that  standard efficiency bound arguments are insufficient in this context, and instead minimax efficiency is the more pertinent benchmark. However, minimax efficiency is not thoroughly understood even for relatively simple causal effect parameters like $\beta$ \citep{robins2017minimax}.  

We commend H{\"a}ggstr{\"o}m again for her interesting proposal of incorporating graph estimation methods in confounder selection, and hope our discussion might help spur future research in this and related areas.

\backmatter

%%%%%% include this section if you wish to acknowledge people,
%%%%%% grant support, etc.

%\section*{Acknowledgements}

%The authors thank anonymous referee for very useful comments that improved the presentation of the paper.
%\vspace*{-8pt}

%%%%%% include this section only if your manuscript refers to supplementary
%%%%%% materials -- see Instructions for Authors at 
%%%%%% http://www.tibs.org/biometrics

\label{lastpage}

\end{document}